\documentclass[aps,prl, reprint,tightenlines,twocolumn,superscriptaddress,showbibnotes,showpacs,tighting]{revtex4-1}

\usepackage{graphicx,epstopdf}
\usepackage{amsmath}
\usepackage{amssymb}
\usepackage{bbm}
\usepackage{color}
\usepackage{float}
\usepackage{epsfig}
\usepackage{epstopdf}
\usepackage{framed}
\usepackage{hhline}
\usepackage[table]{xcolor}
\usepackage{enumerate}
\usepackage{empheq}

\usepackage[colorlinks=true, citecolor=blue, urlcolor=blue ]{hyperref}
\input{epsf}
\usepackage{bbm, dsfont}

\newcommand{\bk}[2]{\langle#1|#2\rangle}

\usepackage{graphicx}
\usepackage{amsmath, amsthm, amssymb}
\usepackage{bm}
\usepackage{epsfig,color}

\newcommand{\tr}{\text{tr}}
\newcommand{\ket}[1]{| #1 \rangle}
\newcommand{\bra}[1]{\langle #1|}

\newcommand{\be}{\begin{equation}}
\newcommand{\ee}{\end{equation}}
\newcommand{\bea}{\begin{eqnarray}}
\newcommand{\eea}{\end{eqnarray}}
\newcommand{\bes}{\begin{equation*}}
\newcommand{\ees}{\end{equation*}}
\newcommand{\beas}{\begin{eqnarray*}}
\newcommand{\eeas}{\end{eqnarray*}}

\newcommand{\ketbra}[1]{\ket{#1}\!\bra{#1}}

\def\tr{\mathrm{tr}}

\newtheorem*{thm*}{Theorem}

\newtheorem*{lem*}{Lemma}

\newtheorem*{lipschitzLem*}{Lemma \ref{lipschitz}}
\newtheorem*{lipschitzCubeLem*}{Lemma \ref{lipschitzCube}}
\newtheorem*{pgmNearlyOptimalThm*}{Theorem \ref{pgmNearlyOptimal}}

\begin{document}

\title{Detecting Entanglement can be More Effective with Inequivalent Mutually Unbiased Bases}

\author{B. C. Hiesmayr}
\email{Beatrix.Hiesmayr@univie.ac.at}
\affiliation{University of Vienna, Boltzmanngasse 5, 1090 Vienna,}

\author{D. McNulty}
\email{dam91@aber.ac.uk}
\affiliation{ Department of Mathematics,  Aberystwyth University, Aberystwyth, Wales, U.K.,}

\author{S. Baek }
\affiliation{Department of Applied Mathematics, Hanyang University (ERICA), 55 Hanyangdaehak-ro, Ansan, Gyeonggi-do, 426-791, Korea,}

\author{S. Singha Roy }
\affiliation{ Instituto de F{\'i}sica Te{\'o}rica UAM/CSIC, C/ Nicol{\'a}s Cabrera 13-15, Cantoblanco, 28049 Madrid, Spain,}


\author{J. Bae }
\email{joonwoo.bae@kaist.ac.kr}
\affiliation{School of Electrical Engineering, Korea Advanced Institute of Science and Technology (KAIST), 291 Daehak-ro Yuseong-gu, Daejeon 34141 Republic of Korea,}

\author{D. Chru\'sci\'nski}
\email{darch@fizyka.umk.pl}
\affiliation{Institute of Physics, Faculty of Physics, Astronomy, and Informatics, Nicolaus Copernicus University, Grudziadzka 5, 87-100 Torun, Poland.}


\begin{abstract}
Mutually unbiased bases (MUBs) provide a standard tool in the verification of quantum states, especially when harnessing a complete set for optimal quantum state tomography. In this work, we investigate the detection of entanglement via inequivalent sets of MUBs, with a particular focus on unextendible MUBs. These are bases for which an additional unbiased basis cannot be constructed and, consequently, are unsuitable for quantum state verification. Here, we show that unextendible MUBs, as well as other inequivalent sets in higher dimensions, can be more effective in the verification of entanglement. Furthermore, we provide an efficient and systematic method to search for inequivalent MUBs and show that such sets occur regularly within the Heisenberg-Weyl MUBs, as the dimension increases. Our findings are particularly useful for experimentalists since adding optimal MUBs to an experimental setup enables a step-by-step approach to detect a larger class of entangled states.
\end{abstract}


\maketitle


\emph{Introduction.}---Quantum entanglement is one of the key ingredients responsible for many of the recent advances in quantum technologies, however, the detection of this fundamental property, even with full knowledge of the quantum state, is in general a NP-hard problem~\cite{Gurvits}. In this contribution we exploit an experimentally feasible protocol to detect entanglement in which the subsequent addition of measurement settings detect a larger class of entangled states. The essential feature of this scheme is for the measurement settings to exhibit complementarity, namely, that the measurements form a set of mutually unbiased bases (MUBs), i.e. the overlap of any pair of vectors from different bases is constant \cite{schwinger}. Physically, this means that exact knowledge of the measurement outcome of one observable implies maximal uncertainty in the other.

Utilising this property, MUBs play an important role in many information processing tasks, such as quantum state tomography \cite{wootters,ivanovic}, quantum key distribution \cite{bennett}, signal processing \cite{sarwate79}, and quantum error correction \cite{calderbank}, to name just a few. Unfortunately, the existence of maximal sets of these highly symmetric bases is a difficult unresolved question: an equivalent formulation of the problem in terms of orthogonal decompositions of the algebra $\mathfrak{sl}_{d}(\mathbb{C})$ dates back almost forty years \cite{kostrikin81,klapp}. Whilst it is known that maximal sets of $d+1$ MUBs in $\mathbb{C}^d$ exist when $d$ is a prime or prime-power, e.g. by a construction based on the Heisenberg-Weyl group \cite{bandyo}, it is conjectured that fewer MUBs exist in all other dimensions \cite{boykin07,durt10}. The classification of subsets of MUBs is an open problem for $d>5$, and a rich structure of inequivalent MUBs (up to unitary transformations) exists \cite{kantor12}. Of particular relevance to this work is the property of unextendibility, i.e. sets of MUBs which cannot be completed to a maximal set.

One striking feature of the entanglement witness, first introduced in~\cite{MUBHiesmayr}, is that the value of the upper bound (which is violated by entanglement) depends only on the \emph{number} of MUBs and not on the \emph{choice} between inequivalent sets. For example, the experimenter is free to use extendible or unextendible MUBs. The protocol has since been applied to two photons entangled in their orbital angular momentum, and bound entangled (PPT-entangled) states have been verified experimentally for the first time~\cite{BoundExpHiesmayr,BoundHiesmayr}. Further modifications of the witness have also been considered, e.g. with $2$-designs and orthogonal rotations \cite{kdesign,darek18,Karol2}, as well as in applications to other scenarios~\cite{BoundNuclear,CVexp}.

Recently, entanglement witnesses were also shown to have a lower bound~\cite{witness2:0,KalevBae}, which often turns out to be non-trivial, i.e. entanglement is detectable. In this contribution, we compute the lower bounds of the MUB-witness and reveal that, in contrast to the upper bound, the values depend strongly on the \emph{choice} of MUBs, as well as the number of measurements. One consequence of this sensitivity,  as we will see, is that unextendible MUBs can be more effective at entanglement detection than extendible ones. Furthermore, this operational distinction between MUBs reveals the existence of inequivalent sets and hence the witness provides a way to classify these sets. We use this criterion to establish inequivalences within the Heisenberg-Weyl MUBs when $d=5,7,9$, as well as for some continuous families and unextendible sets.

This contribution is perhaps the first application of MUBs in which unextendible sets may be the preferred measurement choice over Heisenberg-Weyl subsets. In fact, most applications exhibit no preference for a particular subset of MUBs, and hence the physical differences between inequivalent sets have not been fully realised.  Our results complement some recent observations on measurement incompatibility \cite{designolle19} and quantum random access codes (QRAC) \cite{tavakoli,aguilar}. In particular, inequivalent MUBs contain varying degrees of incompatibility when quantified by their noise robustness, meaning that some sets require additional noise to become jointly measurable. Furthermore, measuring different subsets of MUBs in a QRAC protocol reveals ``anomalies''  in the average success probability, which appear to coincide with our different lower bounds of the witness. Here, by considering entanglement witnesses, we provide a physically significant application which exploits these inequivalences.
\\

\emph{Inequivalent and unextendible MUBs.}---Formally, we say a pair of bases in $\mathbb{C}^d$, labelled $\mathcal{B}_k=\{\ket{i_k}\}_{i=1}^{d}$ and $\mathcal{B}_{k'}$, are mutually unbiased iff
$$|\langle i_k|i'_{ k' }\rangle|^2 = \delta_{i,i'} \delta_{k,k'}+(1-\delta_{k,k'})\frac{1}{d}\,,$$
for all $i,i'=0,\ldots,d-1$. For prime dimensions the standard Heisenberg-Weyl group provides the essential building blocks for the construction of a complete set of $d+1$ MUBs, while for prime-powers $d=p^n$ the generalised tensor product Heisengerg-Weyl group (as described in the Appendix) is used. However, in almost all dimensions, these provide only a small subset of all possible cases. In order to classify MUBs, we need to introduce the notion of equivalence classes. Two sets of $m$ MUBs are equivalent,
$\{\mathcal{B}_1 ,\ldots, \mathcal{B}_m\}\sim\{\mathcal{B}'_1 ,\ldots, \mathcal{B}'_m\}$,
if one set can be transformed into the other by a unitary or antiunitary transformation, permutations within (or of) bases, and phase factor multiplications. The task of classification is challenging, with success currently limited to $d\leq 5$~\cite{haagerup,brierley10}. When $d=2,3,5$ the Heisenberg-Weyl MUBs exhaust all possibilities: no inequivalence occurs when $d=2,3$ and only two inequivalent triples appear among all possible subsets in $d=5$. In contrast, $d=4$ yields a one-parameter family of pairs, and a three-parameter family of triples, inequivalent to all Heisenberg-Weyl subsets \cite{kraus}. In higher dimensions $d\geq6$ the situation is complicated and closely related to the (very old  \cite{hadamard,tadej06}) problem of searching for Hadamard matrices. In some instances, constructions of MUBs produce cases which do not extend to complete sets, and are aptly named \textit{unextendible} MUBs \cite{mandayam,thas}. In dimensions $d=p^n$, the first examples appear when $d=4$ \cite{brierley10} and $d=7$ \cite{grassl}, and more generally $d=p^2$ \cite{szanto,jedwab}.

Finally, we fix some notation. It is often convenient to represent MUBs as sets of unitary matrices, where the columns correspond to orthogonal basis vectors. Due to equivalence transformations, it is possible to express a set of $m$ MUBs for $\mathbb{C}^d$ in Hadamard form,
\begin{equation}\nonumber
\{\mathcal{B}_1,\mathcal{B}_2,\ldots,\mathcal{B}_m\}\sim\{I,H_1,\ldots,H_{m-1}\}\,,
\end{equation}
where $I$ is the identity matrix and $H_i$ are $(d\times d)$ complex Hadamard matrices with $|h_{ij}|=1/\sqrt{d}$. Wlog we assume $\mathcal{B}_1\equiv I$ is the standard basis throughout.
\\

\renewcommand{\arraystretch}{1.2}
\begin{table}
\begin{center}
\begin{tabular}{||c|c|c|c|c||c|c|c||}
\hhline{|t=:=:=:=:=:=:=:=t|}
&\cellcolor{blue!50}$d=2$&\cellcolor{blue!50}$d=3$&\multicolumn{2}{c||}{\cellcolor{blue!50}$d=4$}& $\cellcolor{blue!50}d=2$&$\cellcolor{blue!50}d=3$&$\cellcolor{blue!50}d=4$\\
\hhline{||-|-|-|-|-||-|-|-||}
$m$&$\vphantom{\biggl\lbrace} L_{m}$ ~&~ $L_{m}$ ~&~ $L_{m}^{\textrm{ext.}}$ ~&~ $L_{m}^{\textrm{unext.}}$  ~&~ $U_{m}$ ~&~ $U_{m}$
~&~  $U_{m}$    \\
\hhline{|:=====||===:|}
$2$  & $\frac{1}{2}$& $0.211$& $0$  & 0  & $\frac{3}{2}$ & $\frac{4}{3}$ & $\frac{5}{4}$  \\ \hhline{||-|-|-|-|-|-|-|-||}
$3$  &  $1$& $\frac{1}{2}$& $\frac{1}{4}$ & \cellcolor{red!50} $(\frac{1}{4},\frac{1}{2}]$   & $2$    & $\frac{5}{3}$ &  $\frac{6}{4}$ \\ \hhline{||-|-|-|-|-|-|-|-||}
$4$ & & $1$& $\frac{1}{2}$  & $--$          &   &    $2$  & $\frac{7}{4}$\\ \hhline{||-|-|-|-|-|-|-|-||}
$5$ & & & $1$  & $--$  &  &  &  $2$ \\
\hhline{|b:=====:b:===:b|}
\end{tabular}
\end{center}
\caption{Lower ($L_m$) and upper ($U_m$) bounds of the MUB-witness are summarized for $m$ MUBs and $d=2,3,4$. For $d=4$, the unextendible sets lead to a stricter bound.}
\label{tab:MUB}
\end{table}

\renewcommand{\arraystretch}{1.2}
\setlength{\arrayrulewidth}{0.5pt}
\begin{table}
\begin{tabular}{||c|c|| c| c||}
\hhline{|t====t|}
\multicolumn{4}{||c||}{$\cellcolor{blue!50}d=5$}\\
\hhline{|:==:==:|}
m&$\mathcal{O}$ & L  & U   \\ [0.5ex]
\hhline{|:==:==:|}
2& $\mathcal{B}_{1} \mathcal{B}_{2}$   & 0.0297 & $\frac{6}{5}$    \\
\hhline{||-|-||-|-||}
3& $\cellcolor{green!50}\mathcal{B}_{1} \mathcal{B}_{2}  \mathcal{B}_{3} $ [$10$ times]&  $0.2764$  &  $\frac{7}{5}$  \\
\hline
3& $\cellcolor{green!50}\mathcal{B}_{1} \mathcal{B}_{2}  \mathcal{B}_{4} $  [$10$ times] &  $0.1273$ &   $\frac{7}{5}$ \\
\hline
4& $\mathcal{B}_{1} \mathcal{B}_{2}  \mathcal{B}_{3} \mathcal{B}_{4} $  & $0.3350$ &   $\frac{8}{5}$\\
\hline
5& $\mathcal{B}_{1} \mathcal{B}_{2}  \mathcal{B}_{3} \mathcal{B}_{4} \mathcal{B}_{5} $ & $\frac{1}{2}$ &   $\frac{9}{5}$  \\
\hline
6& $\mathcal{B}_{1} \mathcal{B}_{2}  \mathcal{B}_{3} \mathcal{B}_{4} \mathcal{B}_{5}  \mathcal{B}_{6} $ & 1 \ &   2\\
\hhline{|b:==:b:==:b|}
\end{tabular}
\caption{Lower ($L_m$) and upper ($U_m$) bounds of the function $M_m$ for the Heisenberg-Weyl MUBs in $d=5$.}
\label{table:five}\
\end{table}


\emph{Entanglement detection via MUBs.}---In Ref.~\cite{MUBHiesmayr} an experimentally friendly protocol was introduced that can verify the entanglement of a bipartite state $\rho$ on $\mathbb{C}^d\otimes\mathbb{C}^d$, by subsequently measuring a set of MUBs. Both parties, which may be locally separated, and usually called Alice and Bob, each possess the (same) set of $m$ MUBs in $\mathbb{C}^d$. Applying identical projections from each basis, they calculate
\begin{equation}
M_m(\rho):=\sum_{k=1}^m\sum_{i=0}^{d-1}P(i,i|\mathcal{B}_k,\mathcal{B}_k)\,,
\end{equation}
where $P(i,i|\mathcal{B}_k,\mathcal{B}_k)=\tr(\ketbra{i_k}\otimes\ketbra{i_k}\rho)$ denotes the joint probability of Alice and Bob each obtaining outcome $i$ of the basis measurement  $\mathcal{B}_k$, given the state $\rho$. The authors of Ref.~\cite{MUBHiesmayr} showed that the correlation function $M_m$ is bounded above for all separable states by $U_m$ (see Eq. (\ref{eq:upperbound})). This is easily seen when we assume that the source produces the separable state $|i_l i_l\rangle$, with $\ket{i_l}\in\mathcal{B}_l$. Here, the joint probability $P(i,i|\mathcal{B}_l,\mathcal{B}_l)$ is obviously maximal, i.e. $1$, and for any other basis choice $k\not=l$, the joint probabilities result in $\sum_i P(i,i|\mathcal{B}_k,\mathcal{B}_k)=\frac{1}{d}$ due to the unbiasedness condition. The bound follows by exploiting the arithmetic mean and the convexity of separable states. Recently, it was shown in~\cite{witness2:0} that the very definition of an entanglement witness also yields a non-trivial second bound, which for the MUB-witness results in a lower bound. Thus, the quantity $M_m$ has two bounds,
\begin{eqnarray}\label{MUBwitness}
L_m\;\stackrel{\forall\textrm{ separable states}}{\leq}\;M_m(\rho)\;
\stackrel{\forall\textrm{ separable states}}{\leq}\;U_m
\end{eqnarray}
and any violation of the above inequality detects entanglement of the state $\rho$. The upper bound has the simple form
\begin{equation}\label{eq:upperbound}
U_m=1+\frac{m-1}{d}\,,
\end{equation}
and is independent of the \emph{choice} of MUBs, which implies that only the unbiasedness property is exploited \cite{MUBHiesmayr}. One should also note that although the upper bound is independent of the \emph{ordering} of the MUBs, the results depend on the initial correlation of the considered quantum state $\rho$. This in turn can be compensated by Alice or Bob applying local unitaries to the state. Further note that observing entanglement in any \emph{pure} quantum state via the above inequality only requires measurements in (any) two MUBs. For a complete set, the bounds are $L_{d+1}=1$ and $U_{d+1}=2$, which follow from the 2-design property of MUBs \cite{kdesign}, and the witness is most effective at detecting entangled states.


\emph{Detection with inequivalent MUBs.}---In striking contrast with the upper bound of the function $M_m(\rho)$, the lower bound $L_m$ depends not only on the dimension $d$ and the number of MUBs, $m$, but also on the choice of MUBs. In other words, the bound is highly sensitive to the particular set of MUBs chosen, and therefore inequivalent sets often yield different values of $L_m$. In the following, we analyse this dependence to understand its effect on entanglement detection. We restrict our analysis to dimensions $d<10$ to ensure that the numerical optimisations of the bounds are reliable, where we exploit the composite parameterization of unitaries introduced by~\cite{CPHiesmayr}.  Let us first formulate three main observations in the form of theorems, which we will further discuss below:\\
\\
{\textbf{Theorem A:} If sets of $m$ MUBs have different lower bounds $L_m=\min_{\forall\rho_\textrm{SEP}}  M_m(\rho_\textrm{SEP})$, they are inequivalent.\\
\\
{\textbf{Theorem B:} Unextendible MUBs can result in a stricter lower bound $L_m$ and may therefore detect a larger set of entangled states.\\
\\
{\textbf{Theorem C:} Unextendible sets of MUBs may detect a larger set of entangled states via both the upper and lower bounds, $U_m$ and $L_m$.\\


\emph{Unextendible MUBs in $d=4$.}---The first inequivalent sets of MUBs appear in dimension four, and provide a simple example of \textit{Theorem A}. There exists a one-parameter family of inequivalent pairs $\{I,F(x)\}$ and a three-parameter family of triples $\{I,F(x),H(y,z)\}$ where $H_1=F(x)$ is the Fourier family and $H_2=H(y,z)$ is a family of Hadamard matrices defined in the Appendix with $x,y,z\in[0,\pi]$. These cover all pairs and triples in $d=4$ and coincide with the Heisenberg-Weyl case when $x=y=z=\frac{\pi}{2}$ (for which it extends to a complete set). The triple is unextendible for all other parameter choices. As summarized in Table~\ref{tab:MUB} the upper bound is calculated by Eq. (\ref{eq:upperbound}), but the lower bound $L_3$ depends strongly on the choice of triple. For the (extendible) Heisenberg-Weyl triple, we find $L_3^{\text{ext}}=1/4$, whereas $L^{\text{unext}}_3\in(1/4,1/2]$ for all unextendible sets. The strongest possible bound $L^{\text{unext}}_3=1/2$ is reached when $x=\pi/2$ and $y=z=0$, and is achieved for any pure separable state $\rho=\ketbra{a,b}$ with $\bk{a}{b}$=0. In contrast, the Heisenberg-Weyl triple saturates the lower bound only for particular separable states, e.g. $\frac{1}{\sqrt{2}}(|0\rangle-|2\rangle)\otimes\frac{1}{\sqrt{2}}(|1\rangle+|3\rangle)$. We note that $L_2=0$ for all inequivalent pairs.
\\

\emph{Unextendible MUBs are more efficient.}---Let us now study whether unextendible MUBs are more efficient at detecting entanglement (\textit{Theorems B} and \textit{C}), which is suggested by the stricter lower bounds. However, this need not be the case since the witness itself is also altered and may act differently on any given state.

Let us consider a state from the magic simplex~\cite{MagicSimplex1,MagicSimplex2,MagicSimplex3}, e.g. $\rho_{\alpha,\beta}=(1-\alpha-\beta) \mathbbm{1}/d^2+\alpha P_{0,0}+\beta P_{0,1}$ (it applies for any two Bell states), where the entanglement properties are also fully known. Here $P_{0,0}=\frac{1}{d}\sum_{s,t=0}^{d-1} |ss\rangle\langle tt|$ denotes a Bell state and any other Bell state $P_{k,l}$ can be obtained by locally applying in one subsystem a unitary Wely operator $W_{(k,l)}=\sum_{j=0}^{d-1} \omega_d^{j k} |j\rangle\langle j+l|$ with $\omega_d=e^{\frac{2 \pi i}{d}}$. The results for $d=4$ and $\alpha=\beta$ are summarized in Fig.~\ref{fig:lowerbound1} and show the unextendible sets are more efficient in detecting entanglement, as described in \textit{Theorems B \& C}.

A second example is provided by Werner states $\rho_W$, which are invariant under any local unitary $U\otimes U$ \cite{werner89}. The lower bound of (\ref{MUBwitness}) detects all entangled Werner states for a complete set of MUBs, and becomes less effective as the number of measurements is reduced. Due to  $U\otimes U$ invariance, the quantity $M_m(\rho_W)$ is independent of the choice of MUBs, therefore the value of the bound plays a fundamental role in its effectiveness. For three MUBs, the unextendible triple with the strictest lower bound $L_3^{\text{unext}}=1/2$ detects the largest set of entangled states.

\emph{Inequivalent subsets.}---We now analyse the lower bound $L_m$ for all subsets of Heisenberg-Weyl MUBs when $d=5,\dots,9$, to search for inequivalent sets.

\emph{Dimension $d=5$}:  Here, a full classification is already known~\cite{haagerup,brierley10}, and our search recovers all equivalence classes. All subsets of equal cardinality are equivalent except for triples, which can be grouped into one of two classes:
$\{\mathcal{B}_1,\mathcal{B}_2,\mathcal{B}_3\} \nsim \{\mathcal{B}_1,\mathcal{B}_2,\mathcal{B}_4\}\,.$
Computing the lower bound $L_3$ over all ${6\choose 3}=20$ triples, we find two different bounds, as summarized in Table~\ref{table:five}. No other inequivalent sets appear from the $2^6$ permutations, in agreement with previous results.

\begin{figure}
\begin{center}
\includegraphics[width=0.5\textwidth]{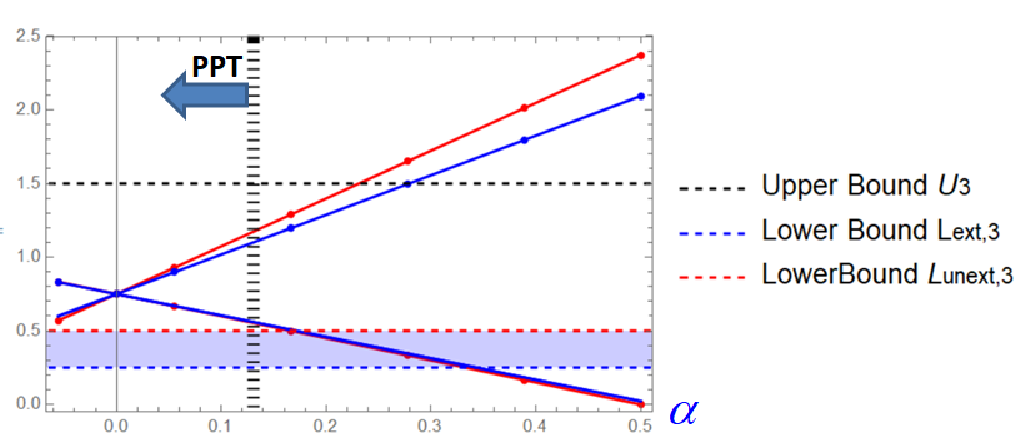}
\caption{The graph shows entanglement detection of extendible and unextendible triples of MUBs in dimension $d=4$, for the family of magic states $\rho(\alpha,\alpha)$. All states on the right hand side of the vertical line are known to be entangled as detected by the $PPT$-criterion. The dots represent the values of $M_3(\rho)$ optimized over local unitarities and exploiting the $U\otimes U^*$ symmetry in the case of unextendible (red) and extendible (blue) sets of MUBs. These provide examples for \textit{Theorems B} and \textit{C}.}
\label{fig:lowerbound1}
\end{center}
\end{figure}

\emph{Dimension $d=7$}: Searching over all subsets, we find only two inequivalent quadruplets
$$\mathcal{Q}_1:=\{\mathcal{B}_1,\mathcal{B}_2,\mathcal{B}_3,\mathcal{B}_4\} \nsim \{\mathcal{B}_1,\mathcal{B}_2,\mathcal{B}_3,\mathcal{B}_5\}:=\mathcal{Q}_2\,.$$
These appear by observing $L_{\mathcal{Q}_1}=0.1514$ and $L_{\mathcal{Q}_2}=0.20101$. There are $\binom{8}{4}=70$ combinations of size four, with $42$ sets achieving the first bound and $28$ the second (higher bound). To explain this distribution, note that no inequivalent triples exist, i.e. any triple is equivalent to $\{\mathcal{B}_1,\mathcal{B}_2,\mathcal{B}_3\}$. Hence, there are only \emph{five} possible extensions to a quadruplet, namely $\{\mathcal{B}_1,\mathcal{B}_2,\mathcal{B}_3,\mathcal{B}_k\}$, $k=4,\ldots,8$. If $k=4,6,8/5,7$ the quadruplets are equivalent to $\mathcal{Q}_1/\mathcal{Q}_2$, therefore the distribution is split as above rather than evenly. We note that two inequivalent quadruplets were also found by analysing the incompatibility content of these subsets~\cite{designolle19} and their success in a QRAC protocol~\cite{aguilar}. This case is also interesting due to the existence of an unextendible triple, $\{\mathcal{B}_1,\mathcal{B}_2,A_7\}$, defined in the Appendix. For this set, the lower bound $L_3=0.0557$ is smaller than the Heisenberg-Weyl bound ($0.0698$). Again, we have examples of the above theorems. The results are summarized in Table~\ref{table:seven}.

\renewcommand{\arraystretch}{1.2}
\setlength{\arrayrulewidth}{0.5pt}
\begin{table}
\begin{tabular}{||c|c|| c| c||}
\hhline{|t====t|}
\multicolumn{4}{||c||}{$\cellcolor{blue!50}d=7$}\\
\hhline{|:==:==:|}
m&$\mathcal{O}$ & L  & U   \\ [0.5ex]
\hhline{|:==:==:|}
2& $\mathcal{B}_{1} \mathcal{B}_{2}$   & 0.0034 & $\frac{8}{7}$    \\
\hhline{||-|-||-|-||}
3& $\mathcal{B}_{1} \mathcal{B}_{2}  \mathcal{B}_{3} $ &  0.0698  &  $\frac{9}{7}$  \\
\hline
3& $\cellcolor{red!50}\mathcal{B}_{1} \mathcal{B}_{2}  \mathcal{A}_7 $   &  $0.0557$ &   $\frac{9}{7}$ \\
\hline
4& $\cellcolor{green!50}\mathcal{B}_{1} \mathcal{B}_{2}  \mathcal{B}_{3} \mathcal{B}_{4} $ [$42$ times]  & $0.1514$ &   $\frac{10}{7}$ \\
\hline
4& $\cellcolor{green!50}\mathcal{B}_{1} \mathcal{B}_{2}  \mathcal{B}_{3} \mathcal{B}_{5} $  [$28$ times]  & $0.20101$ &   $\frac{10}{7}$ \\
\hline
5& $\mathcal{B}_{1} \mathcal{B}_{2}  \mathcal{B}_{3} \mathcal{B}_{4} \mathcal{B}_{5} $ & $0.2750$ &   $\frac{11}{7}$  \\
\hline
6& $\mathcal{B}_{1} \mathcal{B}_{2}  \mathcal{B}_{3} \mathcal{B}_{4} \mathcal{B}_{5} \mathcal{B}_{6}$ & $0.3896$ &   $\frac{12}{7}$  \\
\hline
7& $\mathcal{B}_{1} \mathcal{B}_{2}  \mathcal{B}_{3} \mathcal{B}_{4} \mathcal{B}_{5}  \mathcal{B}_{6} \mathcal{B}_{7}$ & 0.5 \ &   $\frac{13}{7}$\\
\hline
8& $\mathcal{B}_{1} \mathcal{B}_{2}  \mathcal{B}_{3} \mathcal{B}_{4} \mathcal{B}_{5}  \mathcal{B}_{6} \mathcal{B}_{7} \mathcal{B}_{8}$ & 1 \ &   2\\
\hhline{|b:==:b:==:b|}
\end{tabular}
\caption{Lower ($L_m$) and upper ($U_m$) bounds of the function $M_m$ for the Heisenberg-Weyl MUBs and unextendible MUBs in $d=7$.}
\label{table:seven}\
\end{table}
\emph{Dimension $d=8$}: We find a very distinct picture within the Heisenberg-Weyl set, with no inequivalent sets found, in contrast with odd prime and prime-power dimensions. In particular, the bounds are given by $L_{m=2,\dots 4}=0,L_5=\frac{1}{8},L_6=\frac{2}{8},L_7=\frac{3}{8},L_8=\frac{1}{2},L_9={1}$.

\emph{Dimension $d=9$}: Within the Heisenberg-Weyl set we find no inequivalences for $m=2,3$ with $L_{2}=L_{3}=0$, but three inequivalent sets for $m=4$ ($L_4=0,0.077,\frac{1}{6}$) with occurrence $15:180:15$, three inequivalent sets for $m=5$ ($L_5=0.140,0.191,0.198$) with occurrence $90:72:90$, three inequivalent sets for $m=6$ ($L_6=\frac{2}{11},0.259,\frac{1}{3}$) with occurrence $15:180:15$, two inequivalent sets for $m=7$ ($L_7=\frac{1}{3},0.331)$ with occurrence $60:60$ and no inequivalent sets for $m=8,9,10$ ($L_8=0.418,L_9=\frac{1}{2},L_{10}=1$). We note that these results depend heavily on numerical optimizations. The inequivalences are in full agreement with those found in \cite{designolle19,aguilar}, except when $d=9,m=3$, where we detect no inequivalent triples ($L_3=0$) unlike the two cases found in \cite{designolle19} and the anomaly in \cite{aguilar}. Furthermore, we observe that $L_{m}<L_{m+1}$ does not always hold, which is of particular importance for experimental realisations.

\emph{Dimension $d=6$}: This is the first dimension where a maximal set has not been found, and it is conjectured four MUBs do not exist~\cite{zauner}. The Heisenberg-Weyl construction yields only three bases, although many inequivalent pairs and triples exist \cite{karlsson,jaming}. There is only one known unextendible pair, $\{\mathcal{B}_1,S_6\}$, where $S_6$ is the Tao matrix~\cite{tao,brierley09} defined in the Appendix. The lower bound for any pair (extendible and unextendible) results in $L_2=0$. The lower bound for the unextendible Heisenberg-Weyl triple is $L_3=0.1056$.


\emph{Summary \& Outlook.}---We have studied the role that inequivalent MUBs play in the detection of entanglement, as well as providing a method to systematically distinguish inequivalent sets. The dual bounds of the witness exhibit contrasting behaviours, as the choice of MUBs plays a fundamental role in the effectiveness of the lower bound. This is a crucial observation for experimentalists who want to maximise their success in detecting entanglement. Little is known about the physical significance of inequivalent MUBs, and our witness is perhaps the first application where unextendible MUBs are the preferred choice. This leads to questions of whether other applications exist which prioritise one set over another (as is the case for QRACs~\cite{aguilar}), or if inequivalent MUBs have other properties responsible for their varying degrees of usefulness. One such possibility is their incompatibility content, which also distinguishes between equivalence classes of MUBs~\cite{designolle19}. Exploring connections between incompatibility, inequivalent MUBs, and entanglement detection, such as the role incompatibility plays in the effectiveness of entanglement witnesses, may reveal new insights into these topics. Finally, we point out that our findings provide an alternative method to study the structure of the convex set of separable states and subsequently, the rich structure of entanglement.
\\

\emph{Acknowledgements.}---BH acknowledges gratefully the Austrian Science Fund (FWF-P26783). DM has received funding from the European Union's Horizon 2020 research and innovation programme under the Marie Sk\l{}odowska-Curie grant agreement No 663830. JB is supported by the National Research Foundation of Korea (O2N-Q2A 2019M3E4A1080001, NRF-2020K2A9A2A15000061) and the ITRC Program (IITP-2021-2018-0-01402). DC was supported by the National Science Centre project 2018/30/A/ST2/00837. The computational results presented have been achieved in part using the Vienna Scientific Cluster (VSC). We thank M. Grassl for helpful comments and suggestions.

\section*{Appendix}

\subsection*{Heisenberg-Weyl MUBs}

The unitary Heisenberg-Weyl shift and phase operators, acting on $\mathbb{C}^d$, are defined by
\begin{equation}
X=\sum_{j=0}^{d-1} |j+1\rangle\langle j|\,,\qquad Z=\sum_{j=0}^{d-1} \omega_d^j |j\rangle\langle j|\,,
\end{equation}
respectively, where $\omega_d=\exp(2\pi i/d)$. For prime dimensions, $d=p$, the eigenstates of $Z$, $X$, $XZ$, $XZ^{2},\ldots,XZ^{d-1}$ form a complete set of $d+1$ MUBs, and can be written in the concise form
\begin{equation}
\ket{j_{k}}=\frac{1}{\sqrt{p}}\sum_{\ell=0}^{p-1}\omega_p^{(k\ell^{2}+j\ell)}\ket{\ell},
\end{equation}
where $k,j\in\mathbb{F}_{p}$, such that $k$ labels each basis, $j$ the elements within each basis, and $\ket{\ell}$ the elements of the canonical basis. Together with the standard basis, the bases form a set of $p+1$ MUBs. We can also represent this in matrix form, with $B_1= I$, $B_2=F_d$, and
\begin{equation}\label{HWmubs}
\mathcal{B}_{i+1}=D_d^iF_d\,,
\end{equation}
for $i=1,\ldots,d$, where $F_d$ is the $(d\times d)$ Fourier matrix with elements
\begin{equation}
f_{jk}=\frac{1}{\sqrt{d}}\exp(2\pi i(j-1)(k-1)/d)\,,
\end{equation}
and $D_d$ a diagonal matrix. For $d=5,7$, the complete sets are generated using
\begin{equation}
D_5=\text{diag}(1,\omega_5,\omega_5^4,\omega_5^4,\omega_5)
\end{equation}
and
\begin{equation}
D_7=\text{diag}(1, \omega_7, \omega_7^4, \omega_7^2, \omega_7^2, \omega_7^4, \omega_7)\,.
\end{equation}

For $d=p^n$, the operators $X^{k_{1}}Z^{\ell_{1}}\otimes\ldots\otimes X^{k_{n}}Z^{\ell_{n}}$, where $k_{j},\ell_{j}\in\mathbb{F}_{p}$, each acting on the Hilbert space $\mathbb{C}^{p}\otimes\ldots\otimes\mathbb{C}^{p}$, generate the complete set. Partitioning the operators into commuting classes and finding their common eigenvectors yield the $d+1$ bases. If $p$ is odd, the basis elements can be written in the succinct form
\begin{equation}
\ket{j_k}=\frac{1}{\sqrt{d}}\sum_{\ell\in\mathbb{F}_d}\omega_p^{\text{tr}(k\ell^2+j\ell)}\ket{\ell}\,,
\end{equation}
where $\mathbb{F}_d$ is the Galois field, and the trace of $\alpha\in\mathbb{F}_{d}$ is defined as $\text{tr}(\alpha)=\alpha+\alpha^p+\ldots+\alpha^{p^{n-1}}$. Hence, the bases $\mathcal{B}_k$, $k\in\mathbb{F}_d$, together with the standard basis, form the complete set. If $p$ is even, i.e. $d=2^n$, the construction is based on Galois rings \cite{klapp}, with basis elements given by
\begin{equation}
\ket{j_{k}}=\frac{1}{\sqrt{2^{n}}}\sum_{\ell\in\mathcal{T}_{n}}i^{\text{tr}((k+2j)\ell)}\ket{\ell}\,,
\end{equation}
where $k$ and $j$ are elements of the Teichm{\"u}ller set $\mathcal{T}_{n}$. In this case, the trace map $\text{tr}:\text{GR}(4,n)\rightarrow\mathbb{Z}/4\mathbb{Z}$ is defined as $\text{tr}(x)=\sum_{t=0}^{n-1}\sigma^{t}(x)$ where $\sigma$ is the automorphism $\sigma(k+2j)=k^{2}+2j^{2}$.

\subsection*{Unextendible MUBs}
Here, we provide the explicit forms of the unextendible MUBs discussed in the paper. In dimension $d=4$, the three-parameter family of mutually unbiased triples $\{I, F(x), H(y,z)\}$~\cite{kraus}, is given by the one-parameter Fourier family
\begin{eqnarray}\nonumber
F(x)\; =\;  \frac{1}{2} \left(
\begin{array}{cccc}
1 & 1 & 1 & 1 \\
1 & 1 & -1 & -1 \\
1 & - 1 & ie^{ix} & -ie^{ix} \\
1 & -1 & -ie^{ix} & ie^{ix}
\end{array}
\right)\,,\\\nonumber
\end{eqnarray}
 and the two-parameter family of Hadamard matrices
\begin{eqnarray}\nonumber
H(y,z)\;=\; \frac{1}{2} \left(
\begin{array}{cccc}
1 & 1 & 1 & 1 \\
1 & 1 & -1 & -1 \\
-e^{iy} & e^{iy} & e^{iz} & -e^{iz} \\
e^{iy} & -e^{iy} & e^{iz} & -e^{iz}
\end{array}
\right)\;,
\end{eqnarray}
with $x,y,x\in[0,\pi]$. These triples are unextendible unless $x=y=z=\pi/2$, where they coincide with the Heisenberg-Weyl case.

The only unextendible pair in $d=6$ is given by $\{I,S_6\}$, where $S_6$ is the Tao matrix \cite{tao}:
\begin{equation}
S_6 = \frac{1}{\sqrt{6}}
\left(
\begin{array}{cccccc}
1 & 1 & 1 & 1 & 1 & 1 \\
1 & 1 & \omega_3 & \omega_3 & \omega_3^2 & \omega_3^2 \\
1 & \omega_3 & 1 & \omega_3^2 & \omega_3^2 & \omega_3 \\
1 & \omega_3 & \omega_3^2 & 1 & \omega_3 & \omega_3^2 \\
1 & \omega_3^2 & \omega_3^2 & \omega_3 & 1 & \omega_3 \\
1 & \omega_3^2 & \omega_3 & \omega_3^2 & \omega_3 & 1
\end{array}
\right)\,.
\end{equation}

The set $\{I,F_7,A_7\}$ found in \cite{grassl} for $d=7$, is an unextendible triple of MUBs, with $F_7$ the Fourier matrix and setting $\alpha = (\sqrt{-7} - 3)/4$,
\begin{equation}
A_7 = \frac{1}{\sqrt{7}}
\left(
\begin{array}{ccccccc}
\alpha & \alpha & \alpha & 1 & \alpha & 1 & 1 \\
1 & \alpha & \alpha & \alpha & 1 & \alpha & 1 \\
1 & 1 & \alpha & \alpha & \alpha & 1 &\alpha \\
\alpha & 1 & 1 & \alpha & \alpha & \alpha & 1 \\
1 & \alpha & 1 & 1 & \alpha & \alpha & \alpha \\
\alpha & 1 &\alpha & 1 & 1 & \alpha & \alpha \\
\alpha & \alpha & 1 & \alpha & 1 & 1 & \alpha
\end{array}
\right).
\label{A7_UMUB}
\end{equation}

\end{document}